\documentclass[preprintnumbers,article,amsmath,amssymb,floatfix,10pt,prd,onecolumn,
superscriptaddress,nofootinbib]{revtex4-2}
\usepackage{bm}
\usepackage{amsfonts}
\usepackage{latexsym}
\usepackage[latin1]{inputenc}
\usepackage{graphicx}
\usepackage{amsmath}
\usepackage{palatino}
\usepackage{mathpazo}
\usepackage{textcomp}
\usepackage{float}
\usepackage{booktabs}
\usepackage{dcolumn}
\usepackage{ragged2e}
\usepackage{hyperref}
\hypersetup{colorlinks,citecolor=blue}
\hypersetup{colorlinks=true,linkcolor=red,filecolor=magenta,    urlcolor=blue}
\usepackage{amsmath}
\usepackage{xcolor}
\usepackage{orcidlink}
\usepackage{epsfig}
\usepackage{caption}
\usepackage{subcaption}
\usepackage{commath}
\def\jnl@style{\it}
\def\aaref@jnl#1{{\jnl@style#1}}

\def\aaref@jnl#1{{\jnl@style#1}}

\def\aj{\aaref@jnl{AJ}}                   
\def\apj{\aaref@jnl{ApJ}}                 
\def\apjl{\aaref@jnl{ApJ}}                
\def\apjs{\aaref@jnl{ApJS}}               
\def\apss{\aaref@jnl{Ap\&SS}}             
\def\aap{\aaref@jnl{A\&A}}                
\def\aapr{\aaref@jnl{A\&A~Rev.}}          
\def\aaps{\aaref@jnl{A\&AS}}              
\def\mnras{\aaref@jnl{Mon.~Not.~Roy.~Astron.~Soc.}}             
\def\prd{\aaref@jnl{Phys.~Rev.~D}}        
\def\prc{\aaref@jnl{Phys.~Rev.~C}}  
\def\prl{\aaref@jnl{Phys.~Rev.~Lett.}}    
\def\qjras{\aaref@jnl{QJRAS}}             
\def\skytel{\aaref@jnl{S\&T}}             
\def\ssr{\aaref@jnl{Space~Sci.~Rev.}}     
\def\zap{\aaref@jnl{ZAp}}                 
\def\nat{\aaref@jnl{Nature}}              
\def\aplett{\aaref@jnl{Astrophys.~Lett.}} 
\def\apspr{\aaref@jnl{Astrophys.~Space~Phys.~Res.}} 
\def\physrep{\aaref@jnl{Phys.~Rep.}}      
\def\physscr{\aaref@jnl{Phys.~Scr}}       
\def\commat{\aaref@jnl{Comm.~Math.~Phys.}}              
\def\science{\aaref@jnl{Science}}               
\def\cqg{\aaref@jnl{Classical Quant.~Grav.}}            
\def\jpcs{\aaref@jnl{JPCS}}                                     
\def\ijmpd{\aaref@jnl{Int.~J.~Mod.~Phys.~D}}                    
\def\grg{\aaref@jnl{Gen.~Relat.~Gravit.}}               
\def\rpp{\aaref@jnl{Rep.~Prog.~Phys.}}          
\def\npa{\aaref@jnl{Nucl.~Phys.~A}}        
\def\lrr{\aaref@jnl{Living Rev.~Rel.}}                   
\def\jcap{\aaref@jnl{J.~Cosmology Astropart.~Phys.}}    
\def\rmp{\aaref@jnl{Rev.~Mod.~Phys.}}   
\def\epjc{\aaref@jnl{Eur.~Phys.~J.~C}} 
\def\plb{\aaref@jnl{~Phy.~Lett.~B}} 
\def\mpla{\aaref@jnl{Mod.~Phy.~Lett.~A}} 
\def\arxiv{\aaref@jnl{arxiv.org}}

\allowdisplaybreaks[1]

\begin{document}
\color{black}       
\title{The dynamics of matter bounce cosmology in Weyl-type $f(Q,T)$ gravity}

\author{A. Zhadyranova} 
\email[Email: ]{a.a.zhadyranova@gmail.com}
\affiliation{L. N. Gumilyov Eurasian National University, Astana 010008,
Kazakhstan.}

\author{M. Koussour\orcidlink{0000-0002-4188-0572}}
\email[Email: ]{pr.mouhssine@gmail.com}
\affiliation{Department of Physics, University of Hassan II Casablanca, Morocco.} 

\author{S. Bekkhozhayev}
\email[Email: ]{s.o.bekkhozhayev@gmail.com}
\affiliation{L. N. Gumilyov Eurasian National University, Astana 010008,
Kazakhstan.}

%

\begin{abstract}
In this work, we investigate the dynamics of bouncing cosmologies within the framework of Weyl-type $f(Q,T)$ gravity. Here, $Q$ represents the non-metricity of the space-time, is determined by the vector field $w_\mu$, while $T$ represents the trace of the matter energy-momentum tensor. Our objective is to explore the feasibility of avoiding the Big Bang singularity by implementing a matter-bounce cosmology. To achieve this, we consider a specific model with the functional form $f(Q,T)=\alpha Q+\frac{\beta }{6\kappa ^2}T$, where $\alpha$ and $\beta$ are model parameters. We analyze the dynamical parameters associated with this model and examine the influence of the Weyl-type $f(Q,T)$ theory on these parameters. Moreover, we assess the stability of the proposed model to ensure its viability as a cosmological scenario. Through our analysis, we aim to gain insights into the potential implications and consequences of Weyl-type $f(Q,T)$ gravity for bouncing scenarios, contributing to our understanding of alternative gravitational theories in the context of cosmology.

\end{abstract}
\date{\today}
\maketitle

\section{Introduction}
\label{sec1}

In recent years, modified theories of gravity (MTsG) have emerged as viable extensions of Einstein's general relativity (GR) \cite{Clifton}, driven by the need to explain the latest astrophysical observations related to dark energy (DE) and dark matter (DM) \cite{Riess/1998,Riess/2004,Perlmutter/1999,Koivisto/2006, Daniel/2008,Komatsu/2011,Huang/2006}. While GR has provided valuable insights at small scales, studying its behavior on large scales poses significant challenges \cite{Peebles/2003,Sahni/2000}. Therefore, exploring and testing MTsG against observational data has become crucial for understanding the validity and implications of these theories. Among the various MTsG, two prominent frameworks extensively investigated are $f(R)$ gravity and $f(T)$ gravity, which offer geometrical descriptions of gravity. In $f(R)$ theories, the traditional Einstein-Hilbert action is modified by replacing the scalar curvature $R$ with an arbitrary function $f(R)$. This modification introduces a Levi-Civita connection, leading to a space-time with zero torsion and non-metricity but with non-vanishing curvature \cite{L.A.,SA,S.N.-2}. On the other hand, $f(T)$ gravity is an extension of the TEGR (teleparallel equivalent of GR), where the torsion scalar $T$, constructed from the Weitzenbock connection, replaces the curvature. Consequently, the space-time described by $f(T)$ gravity exhibits zero curvature and non-metricity but non-vanishing torsion \cite{R.F.,E.V.,X.R.}. Recently, a novel approach to gravitational theories known as STG (symmetric teleparallel gravity) or $f(Q)$ theory has been proposed by authors \cite{fQ1,fQ2}. This theory offers a fresh geometric interpretation of space-time by incorporating the non-metricity scalar $Q$ as a fundamental quantity. In STG, the affine connection is utilized, leading to the vanishing of curvature and torsion. The inclusion of non-metricity in gravitational theories is motivated by various mathematical and physical considerations. One such motivation arises from the geometric interpretation of the non-metricity scalar $Q$ of the metric tensor. This non-metricity describes the variation in the length of a vector under parallel transport, providing valuable insights into the geometric properties of space-time \cite{fQ3}. 

A significant extension of the $f(Q)$ theory has emerged in the form of $f(Q, T)$ gravity \cite{fQT1}. This modified theory introduces a non-minimal coupling between the gravitational action and the matter sector by replacing the conventional Lagrangian with an arbitrary function that depends on both $Q$ and $T$ (the trace of the matter energy-momentum tensor $T_{\mu\nu}$). The inclusion of $T$ dependence in these models can be attributed to various factors, such as the presence of exotic imperfect fluids and certain quantum effects. This coupling induces a non-zero covariant divergence of $T_{\mu\nu}$, resulting in the deviation of test particles from geodesic paths \cite{Harko2011}. Consequently, these models have been successful in explaining the late-time cosmic accelerated expansion of the Universe \cite{QT2,QT3}. Numerous researchers, such as Najera and Fajardo \cite{QT4}, Shiravand et al. \cite{QT5}, Bourakadi et al. \cite{QT6}, have conducted thorough investigations into various cosmological concepts within the framework of $f(Q, T)$ theory. Arora et al. \cite{QT7} obtained exact solutions for the Friedmann-Lemaitre-Robertson-Walker (FLRW) cosmological model filled with a perfect fluid matter within the framework of the $f(Q, T)$ theory. Similarly, T. H. Loo \cite{QT8} studied the properties and evolution of the Bianchi type-I cosmological model with a perfect fluid in the context of the $f(Q, T)$ theory. Furthermore, Tayde et al. \cite{QT9} specifically concentrated on the modeling of static wormholes within the framework of the $f(Q, T)$ theory. This study explores the matter bounce cosmology within a specific formulation of $f(Q, T)$ gravity, which highlights the non-minimal coupling between $Q$ and $T$. The investigation is conducted within the framework of proper Weyl geometry, where a particular expression for the non-metricity $Q$ is adopted. This expression is derived from the non-conservation of the metric tensor's divergence, represented as $\nabla _{\mu }g_{\alpha \beta
}=-w_{\mu }g_{\alpha \beta }$. Through the utilization of this methodology, we express the non-metricity using a vector field $w_{\mu }$ that, when combined with the metric tensor, completely determines the non-metricity. This formulation is commonly known as Weyl-type $f(Q, T)$ gravity \cite{Weyl1}. Previous research conducted by Yang et al. \cite{Weyl2} extensively explored various aspects of Weyl-type $f(Q, T)$ gravity. Their investigations focused on topics such as geodesic deviation, the Raychaudhuri equation, the Newtonian limit, and tidal forces within this particular framework. Additionally, Koussour \cite{Weyl3} introduced a model-independent approach in Weyl-type $f(Q, T)$ gravity to examine the phenomenon of crossing the phantom divide line. The author specifically studied the behavior of the crossing using the function $f(Q, T)=\alpha Q+\frac{\beta }{6\kappa ^2}T$, where $\alpha$ and $\beta$ are model parameters.

The matter bounce theory, also known as bouncing cosmology, serves as an alternative explanation for various cosmological phenomena, including inflation, while also addressing the singularity problem associated with the standard big bang cosmology \cite{deHaro:2012xj, Moriconi:2016egx, Wang:2003yr, Ijjas:2016tpn, Battefeld:2014uga}. One of the key motivations behind the matter bounce theory is to provide an alternative understanding of the inflationary epoch, a period of rapid expansion in the early Universe. Instead of assuming an initial singularity followed by a sudden expansion, bouncing cosmology proposes that the Universe goes through cycles of contraction and expansion. During the contraction phase, the Universe becomes highly dense and hot, but instead of collapsing into a singularity, it reaches a minimum size and bounces back, initiating an expansion phase. This cyclic nature of the Universe allows for the possibility of inflation to occur within the framework of the matter bounce theory. Furthermore, the matter bounce theory offers a solution to the singularity problem inherent in the big bang cosmology. Instead of postulating the existence of a singularity where the laws of physics break down, the bouncing cosmology suggests that the Universe never experiences a singularity. By undergoing cycles of contraction and expansion, the Universe avoids the need for physics to be undefined at a singular point. This provides a more complete and self-consistent description of the Universe's evolution. Various types of bouncing cosmologies, such as the ekpyrotic bounce and super bounce, have attracted considerable attention within the framework of modified gravitational theories. In standard big bang cosmology, several necessary conditions for the viability of a successful bouncing DE model have been identified \cite{Ijjas:2016tpn}. Firstly, in the vicinity of a bouncing point, the null energy condition, which is similar to $ \dot{H} = -4\pi G \rho (1+\omega) > 0 $, is violated in an FLRW space-time. Secondly, during the contracting phase of the Universe, the scale factor $a(t)$ decreases with cosmic time ($\dot{a}(t)<0 $), and the Hubble parameter $H(t)$ is negative. Conversely, during the expanding phase, the scale factor $a(t)$ increases with cosmic time ($ \dot{a}(t)>0 $), and the Hubble parameter $H(t)$ is positive. At the bouncing point, the scale factor and the Hubble parameter attain their minimum values, namely $ \dot{a}(t)=0 $ and $ H(t)=0 $. Lastly, a crucial criterion for the quintom model is that the EoS parameter $ \omega $ must cross the quintom line $ \omega=-1 $ \cite{Feng:2004ad}. Researchers have extensively explored bounce scenarios in alternative theories of gravity, including $f(R)$, $f(T)$, and $f(G)$ gravity theories \cite{BounceR, BounceT, BounceG,Agrawal0}. A comprehensive review on the subject can be found in \cite{BounceRev}. Recently, a notable contribution by Agrawal et al. \cite{Agrawal1,Agrawal2} investigated bouncing cosmological models within the framework of $f(Q)$ and $f(Q,T)$ gravity. They employed the reconstruction method to constrain the functional form of $f(Q,T)$ and to explore the implications of such models. 

The structure of the paper is as follows: Sec. \ref{sec2} provides an overview of the fundamental framework of the Weyl-type $f(Q,T)$ gravity. In Sec. \ref{sec3}, the cosmological model is established, incorporating a bouncing scale factor and deriving the relevant physical parameters. Furthermore, various dynamical parameters are obtained within the context of the bouncing scenario. The energy conditions of the model are analyzed in this section as well. To assess the stability of the model, a stability analysis is conducted. Finally, Sec. \ref{sec4} concludes the paper by presenting a summary of the findings and offering concluding remarks.

\section{Weyl-type $f(Q,T)$ gravity and field equations}
\label{sec2}

The Weyl-type $f(Q,T)$ theory of gravity represents a geometric extension of symmetric teleparallel gravity. The action for this extended gravitational theory can be expressed as \cite{Weyl1} 
\begin{equation}
S=\int \sqrt{-g}d^{4}x \left[ \kappa ^{2}f(Q,T)-\frac{1}{4}W_{\alpha \beta }W^{\alpha \beta
}-\frac{1}{2}m^{2}w_{\alpha }w^{\alpha }+\lambda (R+6\nabla _{\mu }w^{\mu }-6w_{\mu }w^{\mu })+\mathcal{L}_{m}%
\right].
\label{1}
\end{equation}

Here, $\kappa ^{2}=\frac{1}{16\pi G}$, the symbol $m$ corresponds to the mass of the particle that is connected to the vector field $w_{\alpha}$, $W_{\alpha \beta }=\nabla _{\beta }w_{\alpha }-\nabla _{\alpha
}w_{\beta}$ represents the field strength tensor of the vector field, and $\lambda$ is the Lagrange multiplier. The term $\mathcal{L}_{m}$ denotes the Lagrangian associated with the matter. In addition, $f=f(Q,T)$ represents an arbitrary function that depends on both the non-metricity $Q$ and the trace of the matter energy-momentum tensor $T$. In the action formulation, the second term represents the conventional kinetic term, while the third term serves as a mass term for the vector field. It is important to note that $g=det(g_{\mu\nu})$ denotes the determinant of the metric tensor $g_{\mu\nu}$, and the scalar non-metricity $Q$ can be defined as
\begin{equation}
Q\equiv -g^{\alpha \beta }\left( L_{\nu \beta }^{\mu }L_{\beta \mu }^{\nu
}-L_{\nu \mu }^{\mu }L_{\alpha \beta }^{\nu }\right) ,  \label{2}
\end{equation}%
where $L_{\alpha \beta }^{\lambda }$ denotes the tensor of deformation, is defined as
\begin{equation}
L_{\alpha \beta }^{\lambda }=-\frac{1}{2}g^{\lambda \gamma }\left( Q_{\alpha
\gamma \beta }+Q_{\beta \gamma \alpha }-Q_{\gamma \alpha \beta }\right) .
\label{3}
\end{equation}

In the context of Riemannian geometry, the covariant derivative of the metric tensor vanishes, expressed as $\nabla_{\mu} g_{\alpha\beta} = 0$. However, in Weyl geometry, this expression is modified as
\begin{equation}
\overline{Q}_{\mu \alpha \beta }\equiv \overline{\nabla }_{\mu }g_{\alpha
\beta }=\partial _{\mu }g_{\alpha \beta }-\overline{\Gamma }_{\mu \alpha
}^{\rho }g_{\rho \beta }-\overline{\Gamma }_{\mu \beta }^{\rho }g_{\rho
\alpha }=2w_{\mu }g_{\alpha \beta },  \label{4}
\end{equation}%

Here, $\overline{\Gamma }_{\alpha \beta }^{\lambda }\equiv \Gamma _{\alpha
\beta }^{\lambda }+g_{\alpha \beta }w^{\lambda }-\delta _{\alpha }^{\lambda
}w_{\beta }-\delta _{\beta }^{\lambda }w_{\alpha }$ represents the Weyl connection, $\Gamma _{\alpha\beta }^{\lambda }$ are the corresponding Christoffel symbol in terms of the metric
tensor $g_{\alpha \beta }$, $\overline{\nabla }_{\mu }$ denotes the covariant derivative compatible with the Weyl geometry. The expression above reveals the non-vanishing covariant derivative of the metric tensor in Weyl geometry, where the term $2w_{\mu }g_{\alpha \beta }$ accounts for the Weyl vector field's contribution.

Using Eqs. \eqref{2}-\eqref{4}, we can establish the following relationship,
\begin{equation}
Q=-6w^{2}.  \label{5}
\end{equation}

By performing the variation of the action with respect to the vector field, we can obtain the generalized Proca equation,
\begin{equation}\label{EOM1}
    \nabla^\beta W_{\alpha \beta }-(m^2+12\kappa^2 f_Q+12\lambda)w_\alpha=6\nabla_\alpha \lambda.
\end{equation}

Upon comparing Eq. (\ref{EOM1}) with the standard Proca equation, it becomes apparent that the effective dynamical mass of the vector field can be represented as
\begin{equation}\label{discussion1}
m^2_{\rm{eff}}=m^2+12\kappa^2f_Q+12\lambda.
\end{equation}
where 
\begin{equation}
    f_{Q}\equiv \frac{%
\partial f(Q,T)}{\partial Q}. 
\end{equation}

Moreover, the generalized field equation can be derived by variating the action \eqref{1} with respect to the metric tensor,
\begin{multline}
\frac{1}{2}\left( T_{\alpha \beta }+S_{\alpha \beta }\right) -\kappa
^{2}f_{T}\left( T_{\alpha \beta }+\Theta _{\alpha \beta }\right) =-\frac{%
\kappa ^{2}}{2}g_{\alpha \beta }f
-6\kappa ^{2}f_{Q}w_{\alpha }w_{\beta }+\lambda \left( R_{\alpha \beta
}-6w_{\alpha }w_{\beta }+3g_{\alpha \beta }\nabla _{\rho }w^{\rho }\right) 
\\
+3g_{\alpha \beta }w^{\rho }\nabla _{\rho }\lambda -6w_{(\alpha }\nabla
_{\beta )}\lambda +g_{\alpha \beta }\square \lambda -\nabla _{\alpha }\nabla
_{\beta }\lambda ,
\label{7}
\end{multline}%
where 
\begin{equation}
T_{\alpha \beta }\equiv -\frac{2}{\sqrt{-g}}\frac{\delta (\sqrt{-g}\mathcal{L}_{m})}{%
\delta g^{\alpha \beta }},  \label{8}
\end{equation}%
and%
\begin{equation}
f_{T}\equiv \frac{\partial f(Q,T)}{\partial T},  \label{9}
\end{equation}%
respectively. Further, in order to facilitate the analysis, we introduce the quantity $\Theta_{\alpha\beta}$, which is defined as
\begin{equation}
\Theta _{\alpha \beta }\equiv g^{\mu \nu }\frac{\delta T_{\mu \nu }}{\delta
g_{\alpha \beta }}=g_{\alpha \beta }\mathcal{L}_{m}-2T_{\alpha \beta }-2g^{\mu \nu }%
\frac{\delta ^{2}\mathcal{L}_{m}}{\delta g^{\alpha \beta }\delta g^{\mu \nu }}.
\label{10}
\end{equation}

In the provided field equation, $S_{\alpha\beta}$ denotes the rescaled energy-momentum tensor associated with the free Proca field,
\begin{equation}
S_{\alpha \beta }=-\frac{1}{4}g_{\alpha \beta }W_{\rho \sigma }W^{\rho
\sigma }+W_{\alpha \rho }W_{\beta }^{\rho }-\frac{1}{2}m^{2}g_{\alpha \beta
}w_{\rho }w^{\rho }+m^{2}w_{\alpha }w_{\beta }.  \label{11}
\end{equation}

In the Weyl-type $f(Q,T)$ theory, according to \cite{Weyl1}, the divergence of the matter-energy-momentum tensor can be expressed as
\begin{align}\label{div}
    \nabla^\alpha T_{\alpha\beta}=\frac{\kappa^2}{1+2\kappa^2 f_T}\Big [ 2\nabla_\beta(\mathcal{L}_{m}f_T)-f_T\nabla_\beta T-2T_{\alpha\beta}\nabla^\alpha f_T\Big ].
\end{align}

Thus, the equation above demonstrates that within the framework of the Weyl-type $f(Q,T)$ theory, the matter energy-momentum tensor does not exhibit conservation. It is worth emphasizing that when $f_T = 0$, the energy-momentum tensor becomes conserved.

In this study, our analysis is centered on the dynamics of a cosmological model known as the FLRW Universe. The FLRW Universe is described by a metric that exhibits isotropy, homogeneity, and spatial flatness. This metric is widely used to describe the large-scale structure and evolution of the Universe, providing a simplified yet powerful framework for studying cosmological phenomena, which can be considered as 
\begin{equation}
ds^{2}=-dt^{2}+a^{2}(t)\left[ dx^{2}+dy^{2}+dz^{2}\right],
\label{FLRW}
\end{equation}%
In this context, the symbol $t$ represents the cosmic time, while $a(t)$ denotes the scale factor that characterizes the expansion of the Universe. In addition, it is assumed that the vector field can be parameterized as
\begin{equation}
w_{\alpha }=\left[
\psi (t),0,0,0\right].
\end{equation}

Thus, $w^{2}=w_{\alpha }w^{\alpha }=-\psi
^{2}(t)$ and $Q=-6w^{2}=6\psi ^{2}(t)$. 

Also, we investigate the Universe within the framework of a perfect fluid model. This model assumes that the matter content of the Universe can be described by a continuous distribution of matter with properties that resemble those of a fluid. The energy-momentum tensor, which characterizes the distribution of energy and momentum in the fluid, is defined as
\begin{equation}
\label{EMT}
T_{\mu\nu}=\left(\rho+p\right)u_\mu u_\nu+ p g_{\mu\nu}.
\end{equation}

Here, the symbol $\rho$ represents the energy density, while $p$ denotes the isotropic pressure. The 4-velocity of the fluid is denoted by $u^\alpha$ and satisfies the condition $u_\alpha u^\alpha=-1$. Thus, the energy-momentum tensor takes the form $T^\alpha_\beta=diag\left(-\rho,p,p,p\right)$, and the quantity $\Theta^\alpha_\beta=\delta^\alpha_\beta p-2T^\alpha_\beta$ can be expressed as $\Theta^\alpha_\beta=diag\left(2\rho+p,-p,-p,-p\right)$.

In the cosmological scenario, considering the constraint of a flat space-time, we obtain the following generalized Proca equations,
\begin{eqnarray}
\dot{\psi} &=&\dot{H}+2H^{2}+\psi ^{2}-3H\psi ,  \label{17} \\
\dot{\lambda} &=&\left( -\frac{1}{6}m^{2}-2\kappa ^{2}f_{Q}-2\lambda \right)
\psi =-\frac{1}{6}m_{eff}^{2}\psi ,  \label{18} \\
\partial _{i}\lambda &=&0.  \label{19}
\end{eqnarray}

Here, $H(t) = \frac{\dot{a}}{a}$ denotes the Hubble parameter, which characterizes the rate of expansion of the Universe. The dot symbol (.) represents the derivative with respect to cosmic time $t$.

From Eq. (\ref{7}) and incorporating the provided metric (\ref{FLRW}), we derive the generalized Friedmann equations as \cite{Weyl1}, 
\begin{equation}
\kappa ^{2}f_{T}(\rho +p)+\frac{1}{2}\rho =\frac{\kappa ^{2}}{2}f-\left(
6\kappa ^{2}f_{Q}+\frac{1}{4}m^{2}\right) \psi ^{2}-3\lambda (\psi ^{2}-H^{2})-3\dot{\lambda}(\psi -H),  \label{F1}
\end{equation}%
\begin{equation}
-\frac{1}{2}p =\frac{\kappa ^{2}}{2}f+\frac{m^{2}\psi ^{2}}{4}+\lambda
(3\psi ^{2}+3H^{2}+2\dot{H}) +(3\psi +2H)\dot{\lambda}+\ddot{\lambda}.  \label{F2}
\end{equation}

When considering the specific scenario where $f=0$, $\psi=0$, and $\lambda=\kappa^2$, the gravitational action (\ref{1}) simplifies to the standard Hilbert-Einstein action. As a consequence, the generalized equations (\ref{F1}) and (\ref{F2}) are reduced to the Friedmann equations in GR. More precisely, these equations can be expressed as: $3H^2=\frac{\rho}{2\kappa^2}$ and $2\dot{H}=-\frac{(\rho+p)}{2\kappa^2}$, respectively. In this particular limit, the energy density $\rho$ and pressure $p$ conform to the standard framework of GR.

In order to explore the matter bounce cosmology within the framework of Weyl-type $f(Q, T)$ gravity theory, it is necessary to choose a specific functional form for $f(Q, T)$. For our analysis, we assume the functional form \cite{Weyl1}
\begin{equation}
    f(Q,T)=\alpha Q+\frac{\beta }{6\kappa ^2}T
    \label{fQT}
\end{equation}
where $\alpha$ and $\beta$ are model parameters. The model is characterized by three independent parameters: $M^2= m^{2}/\kappa^2$, which represents the mass of the Weyl field, and $\alpha$ and $\beta$, which respectively quantify the strengths of the coupling between Weyl geometry and matter. 
Moreover, the predictions of the proposed Weyl-type $f(Q, T)$ model align with the spatially flat $\Lambda$CDM model. In accordance with the analyses presented in \cite{Koussour}, the free parameters $\alpha$, $\beta$, and $M$ are constrained by fitting them to observational data. The resulting values are found to be $\alpha=-2.20_{-0.98}^{+0.76}$, $\beta=-0.67^{+0.60}_{-0.51}$ and $M=1.7^{+2.2}_{-1.8}$. Therefore, by setting $f_Q=\alpha$ and $f_T=\frac{\beta }{6\kappa ^2}$, the generalized Friedmann equations given by Eqs. (\ref{F1})-(\ref{F2}) can be reformulated as,
\begin{equation}
\label{F11}
\frac{\beta}{12}p-\left(\frac{\beta }{4}+\frac{1}{2}\right) \rho =3 \alpha  \kappa ^2 \psi ^2+3 \kappa ^2 \left(\psi ^2-H^2\right)+\frac{m^2 \psi ^2}{4},
\end{equation}
\begin{equation}
\label{F22}
\frac{\beta }{12}\rho-\left(\frac{\beta }{4}+\frac{1}{2}\right) p=3 \alpha  \kappa ^2 \psi ^2+\kappa ^2 \left(3 H^2+2 \dot{H}+3 \psi ^2\right)+\frac{m^2 \psi ^2}{4}.
\end{equation}

For the sake of simplicity, we make the convenient choice of setting $\lambda = \kappa^2 = 1$ \cite{Weyl1}. By solving the aforementioned equation under the assumption $H(t) = \psi(t)$, we can express the isotropic pressure and energy density in terms of the Hubble parameter as 
\begin{equation}
    \rho=-\frac{3 \left[H^2 \left(12 (2 \alpha  \beta +3 \alpha +\beta )+(2 \beta +3) M^2\right)+4 \beta  \dot{H}\right]}{2 \left(2 \beta ^2+9 \beta +9\right)},
    \label{rho}
\end{equation}
\begin{equation}
    p=-\frac{6 \left(3 H^2+\dot{H}\right)}{2 \beta +3}-\frac{3 \left[H^2 \left(12 (\alpha +1)+M^2\right)+4 \dot{H}\right]}{2 (\beta +3)}.
    \label{p}
\end{equation}

The EoS parameter is defined as the ratio of pressure to energy density ($\omega=\frac{p}{\rho}$). It can be obtained from the expressions of pressure and energy density, and is given by
\begin{equation}
    \omega=\frac{H^2 \left[24 \alpha  \beta +36 (\alpha +\beta +2)+(2 \beta +3) M^2\right]+12 (\beta +2) \dot{H}}{H^2 \left[12 (2 \alpha  \beta +3 \alpha +\beta )+(2 \beta +3) M^2\right]+4 \beta  \dot{H}}.
    \label{EoS}
\end{equation}

In the present study, our objective is to explore a matter bounce cosmology using the framework of Weyl-type $f(Q,T)$ gravity and the formalism outlined in this section. To achieve this, we consider a scale factor that emulates a matter bounce cosmology within the expressions of the dynamical parameters discussed earlier. Our aim is to investigate the impact of the model parameters on the dynamic behavior of these parameters and analyze their implications within the context of the matter bounce cosmology.

\section{Matter bounce cosmology}
\label{sec3}

In order to understand the behavior of the Universe, it is essential to determine the physical and dynamical parameters governing its evolution. These parameters are often expressed in terms of the Hubble parameter, which characterizes the rate of expansion of the Universe. It is worth noting that the traditional inflationary scenario falls short of providing a complete description of the Universe's past history. As an alternative approach to address this limitation, the concept of a matter bounce has been proposed. The matter bounce scenario offers a potential solution to the shortcomings of inflationary models by suggesting an alternative mechanism for the early Universe. Instead of a period of rapid expansion driven by an inflationary field, the matter bounce posits a contracting phase followed by a bounce into an expanding phase. This provides a bridge between the contracting and expanding epochs, allowing for a more comprehensive understanding of the Universe's evolution.

In the context of investigating different bouncing scenarios in cosmology, researchers have explored several models that offer alternatives to the standard inflationary paradigm. These models aim to address the singularity problem and provide non-singular cosmological evolution. One type of bounce scenario is the symmetric bounce, which was initially introduced by Cai et al. \cite{Cai/2012}. This model requires satisfying additional cosmological behaviors to avoid issues related to the Hubble horizon. As time tends to infinity, the scale factor, Hubble parameter, energy density, and pressure diverge in this model. Another model is the super-bounce, proposed by Koehn et al. \cite{Koehn/2014}. It describes a Universe that collapses at a specific bouncing point ($t_b$) and undergoes a rebirth through a Big Bang without encountering a singularity. As the time approaches $t_b$, the scale factor remains constant, while the Hubble parameter, energy density, and pressure diverge. The oscillatory cosmology model, discussed by Novello et al. \cite{Novello/2008}, describes a cyclic Universe that undergoes repeated expansion and contraction. According to this model, as the time approaches $kt_b$ (where $k$ is a non-zero integer), the scale factor approaches a constant value, while the Hubble parameter, energy density, and pressure exhibit divergence. The matter bounce model, derived from loop quantum cosmology and extensively studied by Singh et al. \cite{Singh/2006}, provides an alternative to inflation. In this model, as the time approaches the bouncing point $t_b$, the scale factor, Hubble parameter, energy density, and pressure remain finite. Lastly, there are models addressing past and future singularities, including the big-rip, little-rip, sudden, and big-freeze singularities, as explored by Nojiri et al. \cite{Nojiri/2005}. These models aim to understand the nature of singularities and their implications for the Universe's evolution.

The main focus of this study is to investigate the matter bounce cosmology within the framework of Weyl-type $f(Q,T)$ gravity. In order to delve into this topic, we adopt a specific form for the scale factor of the Universe $a(t)$ as proposed by Odintsov and Oikonomou \cite{Odintsov/2016},
\begin{equation}
    a(t)=\left(\frac{3}{4}\rho_ct^2+1\right)^\frac{n}{3},
    \label{at}
\end{equation}
where $\rho_c$ is a constant that represents the quantum nature of space-time, while the integer $n$ characterizes specific aspects of the analysis. To simplify the discussion, we focus on the cases where $n$ takes the values of 1, 2, and 3 in our study. The corresponding Hubble parameter $H(t)$ can be derived as
\begin{equation}
    H(t)=\frac{\dot{a}}{a}=\dfrac{\frac{n}{2} \rho_{c}t}{(\frac{3}{4}\rho_{c}t^{2}+1)}.
    \label{Ht}
\end{equation}

The deceleration parameter $q(t)$ is a crucial parameter that characterizes the dynamics of the Universe. A positive value of $q$ signifies a decelerating phase, while a negative value indicates an accelerating phase. The deceleration parameter can be calculated as
\begin{equation}
q(t) = -1-\frac{\dot{H}}{H^{2}}=-1-\frac{4-3\rho_{c}t^{2}}{2n\rho_{c}t^{2}}.
\label{qt}
\end{equation}

\begin{figure}
    \centering
    \begin{subfigure}[b]{0.24\textwidth}
        \centering
        \includegraphics[width=\textwidth]{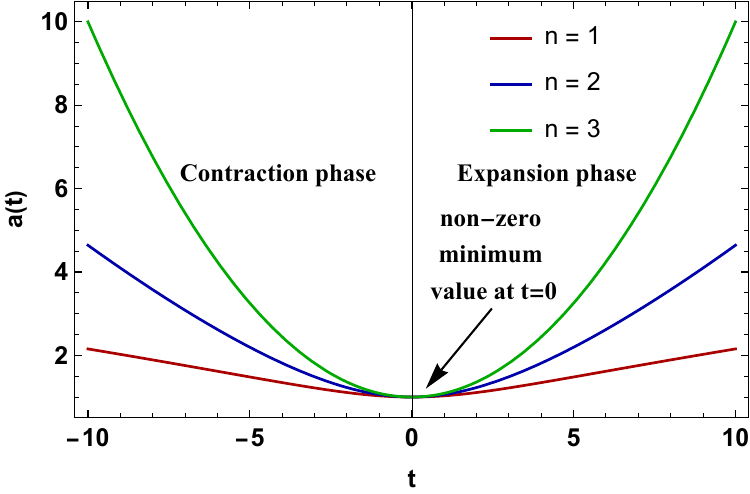}
        \caption{Scale factor}
        \label{F_a}
    \end{subfigure}
    \hfill
    \begin{subfigure}[b]{0.25\textwidth}
        \centering
        \includegraphics[width=\textwidth]{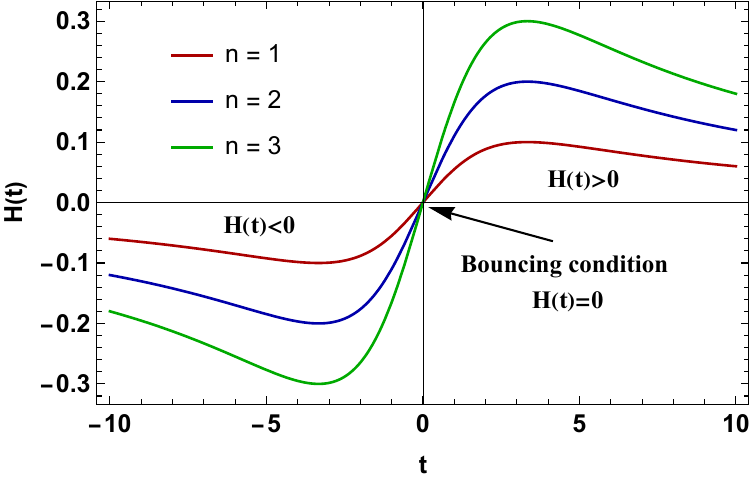}
        \caption{Hubble parameter}
        \label{F_H}
    \end{subfigure}
    \hfill
    \begin{subfigure}[b]{0.24\textwidth}
        \centering
        \includegraphics[width=\textwidth]{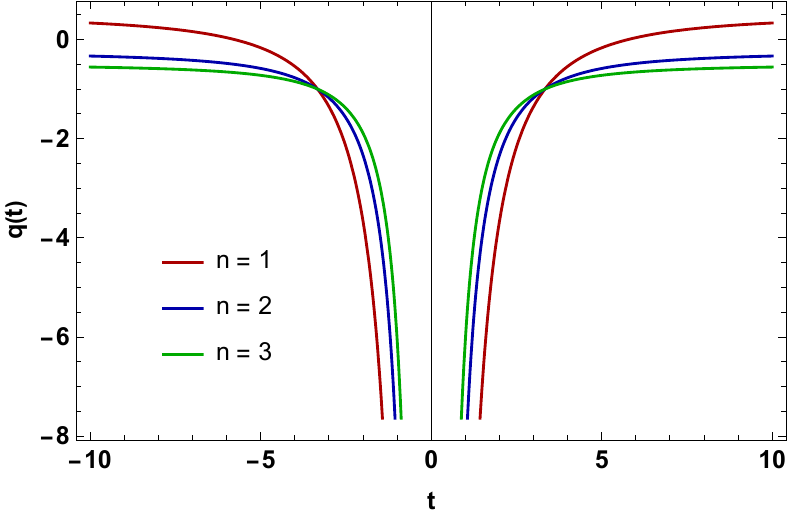}
        \caption{Deceleration parameter}
        \label{F_q}
    \end{subfigure}
    \hfill
    \begin{subfigure}[b]{0.25\textwidth}
        \centering
        \includegraphics[width=\textwidth]{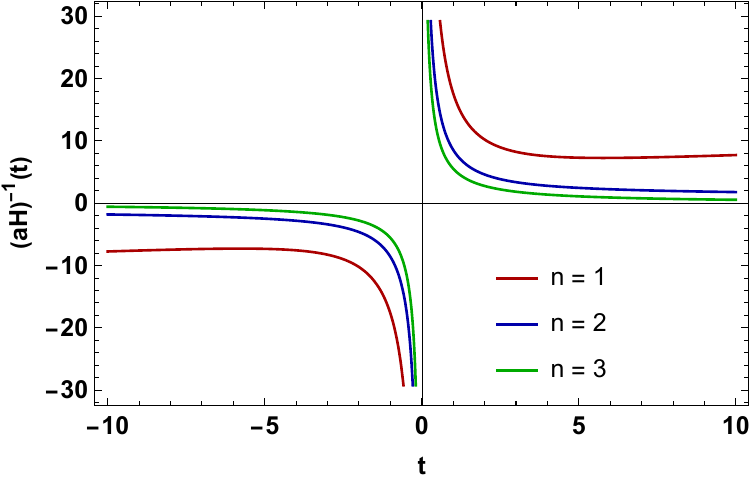}
        \caption{Hubble radius}
        \label{F_RH}
    \end{subfigure}
    \caption{Temporal evolution of cosmological parameters: bouncing scale factor (a), Hubble parameter (b), deceleration parameter (c), and Hubble radius (d).}
    \label{F_par}
\end{figure}

In our analysis, we adopt the standard units of the Hubble, which is expressed in $km/s/Mpc$, and cosmic time, which is measured in $Gyr$. In the matter of bounce cosmology, the behavior of the Universe's contraction and expansion can be understood by studying the Hubble parameter. Fig. \ref{F_par} illustrates the plots of the bouncing scale factor $a(t)$, Hubble parameter $H(t)$, deceleration parameter $q(t)$, and Hubble radius $(aH)^{-1}(t)$ with respect to cosmic time $t$, considering a small value of $\rho_c=0.12$ and three different values of $n$. During the contracting phase of the Universe, the scale factor $a(t)$ monotonically decreases, i.e., $\dot{a(t)} < 0$. On the other hand, during the expanding phase of the Universe, the scale factor $a(t)$ exhibits an increasing pattern, i.e., $\dot{a(t)}>0$. In Fig. \ref{F_a}, we observe that the scale factor of the Universe reaches a non-zero minimum value $a(t)=1$ at the bouncing point $t=t_b$ for $\rho_c=0.01$. This indicates that the volume of the model undergoes a contraction before the bounce and subsequently begins to expand after the bounce. Furthermore, the behavior of the Hubble parameter in Fig. \ref{F_H} provides valuable insights into the dynamics of the cosmic bounce scenario. The transition from negative ($H(t)<0$) to positive values ($H(t)>0$) of $H(t)$ signifies the transition from a contracting phase to an expanding phase of the Universe. The occurrence of the bouncing condition, where $H(t)=0$, indicates a pivotal point where the Universe transitions from contraction to expansion or vice versa. In Fig. \ref{F_q}, the deceleration parameter shows a negative behavior before and after the bouncing point $t_b$. As time progresses away from the bouncing point, the deceleration parameter $q$ continues to increase. In the late-time regime, it converges to $q = -0.5$ for $n=2$ and $n=3$, while for $n=1$, it approaches a positive value. However, near the bouncing point itself, the deceleration parameter exhibits a singularity due to the unique behavior of the bouncing model. Fig. \ref{F_RH} shows that the cosmic Hubble radius experiences a symmetric and monotonic decrease around the bouncing epoch for all values of $n$. Additionally, it asymptotically approaches zero in both the positive and negative time domains. This behavior suggests that the Universe is accelerating in its later stages \cite{Agrawal0,Agrawal1}. Therefore, in these situations, the Hubble horizon shrinks to zero as the cosmic time approaches large values, while only near the bouncing point does the Hubble horizon reach an infinite magnitude.

\subsection{Dynamical Parameters}

To gain a comprehensive understanding of this bouncing model, it is essential to analyze additional factors such as the dynamical parameters. Hence, we need to determine the values of the energy density $\rho$ and isotropic pressure $p$ in the Weyl-type $f(Q,T)$ model. By using Eqs. (\ref{rho}) and (\ref{p}), we can express the energy density and isotropic pressure as
\begin{equation}
    \rho=-\frac{6 n \rho_{c} \left[2 \beta  \left(\rho_{c} t^2 \left(n \left(12 \alpha +M^2+6\right)-3\right)+4\right)+3 n \rho_{c} t^2 \left(12 \alpha +M^2\right)\right]}{\left(2 \beta ^2+9 \beta +9\right) \left(3 \rho_{c} t^2+4\right)^2},
\end{equation}
\begin{equation}
    p=\frac{6 n \rho_{c} \left[-24 (\beta +2)-\left(\rho_{c} t^2 \left(-18 (\beta +2)+2 \beta  n \left(12 \alpha +M^2+18\right)+3 n \left(12 (\alpha +2)+M^2\right)\right)\right)\right]}{(\beta +3) (2 \beta +3) \left(3 \rho_{c} t^2+4\right)^2}.
\end{equation}

Therefore, the EoS parameter becomes,
\begin{equation}
    \omega=\frac{24 (\beta +2)+\rho_{c}t^2 \left[-18 (\beta +2)+2 \beta  n \left(12 \alpha +M^2+18\right)+3 n \left(12 (\alpha +2)+M^2\right)\right]}{8 \beta +\rho_{c} t^2 \left(-6 \beta +(2 \beta +3) M^2 n+12 n (2 \alpha  \beta +3 \alpha +\beta )\right)}.
\end{equation}

\begin{figure}
     \centering
     \begin{subfigure}[b]{0.33\textwidth}
         \centering
         \includegraphics[width=\textwidth]{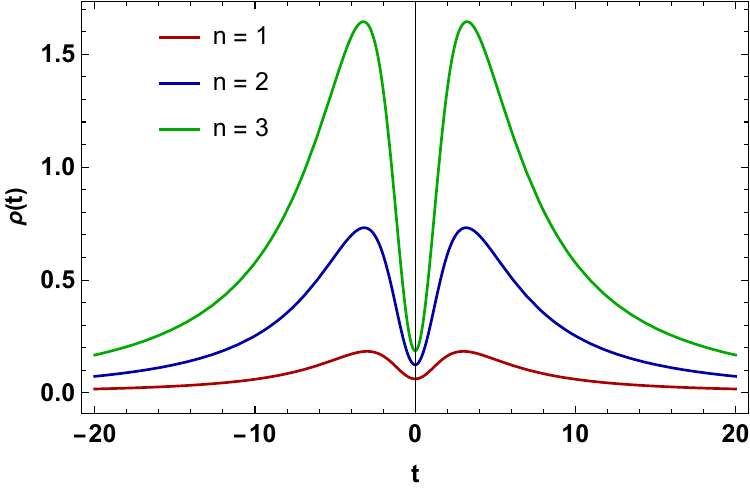}
         \caption{Energy density}
         \label{F_rho}
     \end{subfigure}
     \hfill
     \begin{subfigure}[b]{0.33\textwidth}
         \centering
         \includegraphics[width=\textwidth]{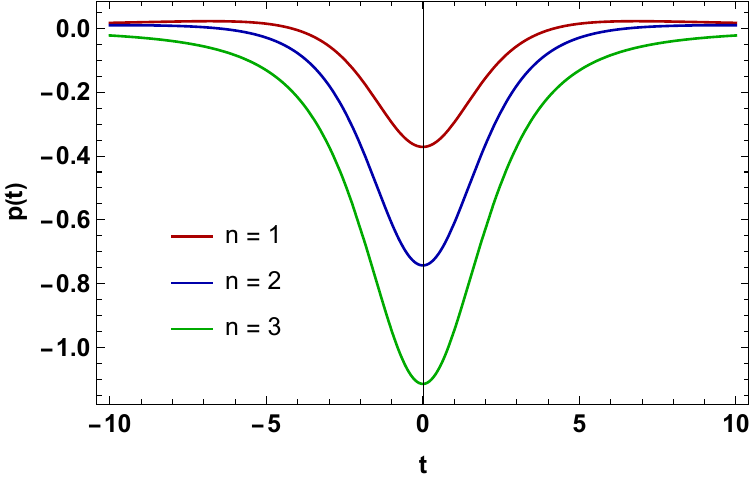}
         \caption{Isotropic pressure}
         \label{F_p}
     \end{subfigure}
     \hfill
     \begin{subfigure}[b]{0.33\textwidth}
         \centering
         \includegraphics[width=\textwidth]{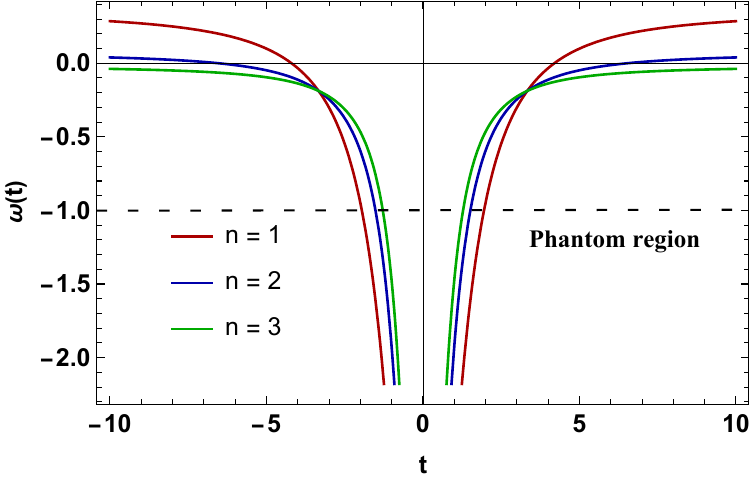}
         \caption{EoS parameter}
         \label{F_EoS}
     \end{subfigure}
        \caption{Temporal evolution of dynamical parameters: energy density (a), isotropic pressure (b), and EoS parameter (c).}
        \label{F_dy}
\end{figure}

Fig. \ref{F_dy} illustrates the behavior of the energy density $\rho$, isotropic pressure $p$, and EoS parameter $\omega$ as a function of cosmic time $t$ for the model parameters $\rho_c=0.12$ and $n=1$, $n=2$, and $n=3$. Fig. \ref{F_rho} shows that the energy density remains positive throughout cosmic evolution. It increases before the bounce epoch as we approach the bouncing point $t_b=0$, forms a trough near the bounce, and then decreases after the bounce region. This behavior is consistent with several works in the literature \cite{Agrawal0,Agrawal1,Agrawal2}. On the other hand, Fig. \ref{F_p} demonstrates that the isotropic pressure maintains a highly negative value throughout the time range before and after the bounce. Furthermore, it is noteworthy that the EoS parameter in this model crosses the quintom line ($\omega=-1$) in the vicinity of the bouncing point $t_b=0$ (see Fig. \ref{F_EoS}). The quintom crossing implies a transition from the phantom era ($\omega<-1$) to the quintessence era ($\omega>-1$), or vice versa. Such a transition is indicative of the model's ability to achieve a smooth and consistent evolution of the Universe, avoiding any singularities or instabilities. Therefore, the observed crossing of the quintom line in the vicinity of the bouncing point further supports the viability and success of the proposed bouncing model \cite{Cai/2012}.

\subsection{Energy Conditions}

The energy conditions impose certain constraints on the combinations of energy density and pressure in a physical system. These conditions provide important insights into the behavior of matter and gravity. The concept of energy conditions emerged from the Raychaudhuri equation \cite{Raychaudhuri}, which describes the evolution of gravitational systems. The energy conditions serve as guidelines that ensure the physical viability of energy-momentum tensors. They establish fundamental principles such as the non-negativity of energy density and the attractive nature of gravity. Specifically, the energy conditions include the null energy condition (NEC), weak energy condition (WEC), dominant energy condition (DEC), and strong energy condition (SEC). For the present model, the energy conditions are derived as \cite{Hawking}
\begin{equation}
NEC\Longleftrightarrow \rho + p=-\frac{12 n \rho_{c} \left[\rho_{c} t^2 \left(n \left(12 \alpha +M^2+12\right)-6\right)+8\right]}{(\beta +3) \left(3 \rho_{c} t^2+4\right)^2}\geq 0,
\end{equation}%
\begin{equation}
WEC\Longleftrightarrow \rho=-\frac{6 n \rho_{c} \left[2 \beta  \left(\rho_{c} t^2 \left(n \left(12 \alpha +M^2+6\right)-3\right)+4\right)+3 n \rho_{c} t^2 \left(12 \alpha +M^2\right)\right]}{\left(2 \beta ^2+9 \beta +9\right) \left(3 \rho_{c} t^2+4\right)^2}\geq 0,
\end{equation}%
\begin{equation}
    DEC\Longleftrightarrow \rho - p=\frac{24 n \rho_{c} \left[3 (2 n-1) \rho_{c} t^2+4\right]}{(2 \beta +3) \left(3 \rho_{c} t^2+4\right)^2}\geq 0,
\end{equation}
\begin{equation}
    SEC\Longleftrightarrow \rho +3p=-\frac{24 n \rho_{c} \left[\beta  \left(\rho_{c} t^2 \left(2 n \left(12 \alpha +M^2+15\right)-15\right)+20\right)+3 \rho_{c} t^2 \left(n \left(12 \alpha +M^2+18\right)-9\right)+36\right]}{(\beta +3) (2 \beta +3) \left(3 \rho_{c} t^2+4\right)^2}\geq 0,
\end{equation}

\begin{figure}
     \centering
     \begin{subfigure}[b]{0.33\textwidth}
         \centering
         \includegraphics[width=\textwidth]{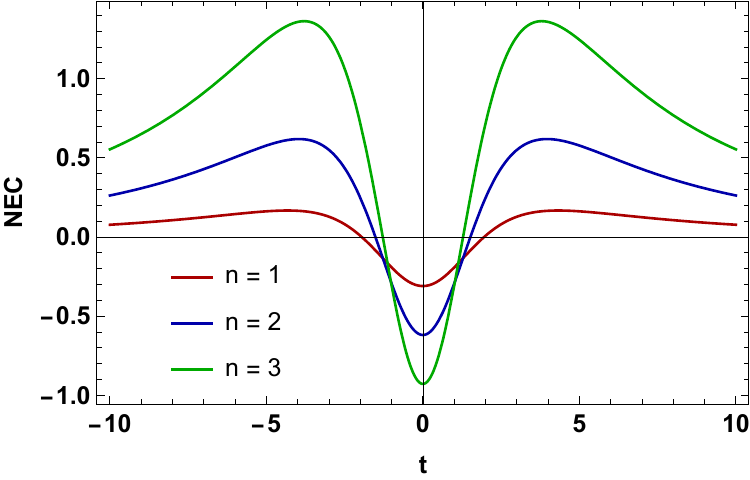}
         \caption{$\rho +p$}
         \label{F_NEC}
     \end{subfigure}
     \hfill
     \begin{subfigure}[b]{0.33\textwidth}
         \centering
         \includegraphics[width=\textwidth]{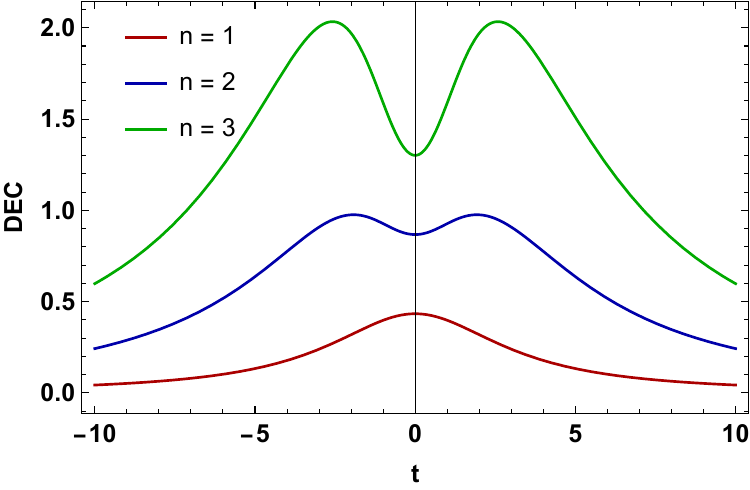}
         \caption{$\rho -p$}
         \label{F_DEC}
     \end{subfigure}
     \hfill
     \begin{subfigure}[b]{0.33\textwidth}
         \centering
         \includegraphics[width=\textwidth]{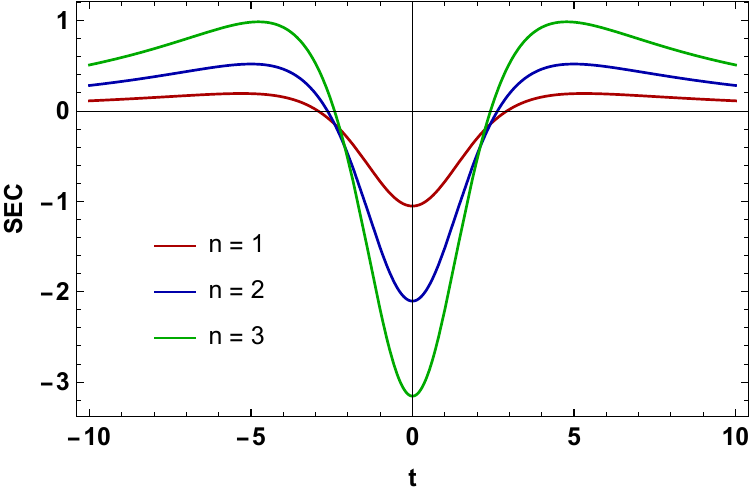}
         \caption{$\rho +3p$}
         \label{F_SEC}
     \end{subfigure}
        \caption{Temporal evolution of energy conditions: NEC (a), DEC (b), and SEC (c).}
        \label{F_ECs}
\end{figure}

These conditions provide important bounds on the relationship between energy density and pressure, ensuring the consistency and physical reasonability of the energy-momentum tensor. 
In order to ensure a successful bouncing model, it is crucial to examine the variations of the NEC and the first derivative of the Hubble parameter $\dot{H}$ with respect to cosmic time i.e. $ \dot{H}=-4 \pi G \rho (1+\omega)>0 $, as depicted in the plots of Fig. \ref{F_ECs} for three different values of $n$. The NEC, given by the condition $\rho+p \geq 0$, serves as an important constraint on the energy density and pressure. In the vicinity of the bouncing point at $t_b=0$, we observe a violation of the NEC (see Fig. \ref{F_NEC}), indicating the presence of exotic matter or energy conditions that deviate from conventional expectations. In addition, the plot reveals that $\dot{H}>0$ in the neighborhood of the bouncing point, satisfying the condition $\omega<-1$ required for a successful bouncing cosmic scenario. This result further strengthens the viability of the proposed bouncing model, as it indicates a positive rate of change of the Hubble parameter and supports a smooth transition between the contraction and expansion phases.

Fig. \ref{F_ECs} provides further insights into the energy conditions in our bouncing model. It is evident from the plot that the DEC (see Fig. \ref{F_DEC}), which states that the energy density $\rho$ must be greater than or equal to the absolute value of the pressure $|p|$, holds true in our model. This condition ensures that the energy density dominates over the pressure, which is an essential criterion for the stability and consistency of the cosmic dynamics. On the other hand, the SEC (see Fig. \ref{F_SEC}), given by the condition $\rho+3p\geq0$, is violated in the vicinity of the bouncing point at $t_b=0$. This violation of the SEC indicates the presence of DE in the Universe, which is consistent with our understanding of the accelerated expansion observed in the late-time Universe.

Therefore, our bouncing model in Weyl-type $f(Q,T)$ gravity satisfies the necessary criteria, including the violation of both NEC and SEC and the fulfillment of DEC. These results demonstrate the presence of DE and the consistent behavior of energy conditions in our model, reinforcing its viability as a successful bouncing cosmological scenario.

\subsection{Stability Analysis}

Here, we will delve into the stability analysis of our cosmological model in Weyl-type $f(Q,T)$ gravity. To assess the stability, we employ a useful tool known as the squared sound speed, denoted by $v_s^2(t)$. This parameter provides crucial information about the propagation of perturbations and the stability of the model. It is defined as the ratio of the perturbation in pressure to the perturbation in energy density \cite{Peebles/2003},
\begin{equation}
v_{s}^{2}\left( t\right) =\frac{dp }{d\rho}=\frac{\overset{.}{p }}{%
\overset{.}{\rho}}.
\end{equation}

The purpose of analyzing the squared sound speed is to determine the stability of the proposed cosmological model. If the squared sound speed $v_{s}^{2}\left( t\right)$ is greater than zero ($v_{s}^{2}\left( t\right) >0$), it indicates that the model is stable. On the other hand, if $v_{s}^{2}\left( t\right)$ is less than zero ($v_{s}^{2}\left( t\right) <0$), it indicates that the model is unstable. In the context of the bounce cosmology proposed in Weyl-type $f(Q,T)$ gravity, the expression for the squared sound speed is given by
\begin{equation}
    v_{s}^{2}\left( t\right) =\frac{n \left(3\rho_{c} t^2-4\right) \left(24 \alpha  \beta +36 (\alpha +\beta +2)+(2 \beta +3) M^2\right)-54 (\beta +2) \left(\rho_{c} t^2-4\right)}{(2 \beta +3) M^2 n \left(3 \rho_{c} t^2-4\right)+12 n (2 \alpha  \beta +3 \alpha +\beta ) \left(3 \rho_{c} t^2-4\right)-18 \beta  \left(\rho_{c} t^2-4\right)}
\end{equation}

\begin{figure}[h]
\centerline{\includegraphics[scale=0.65]{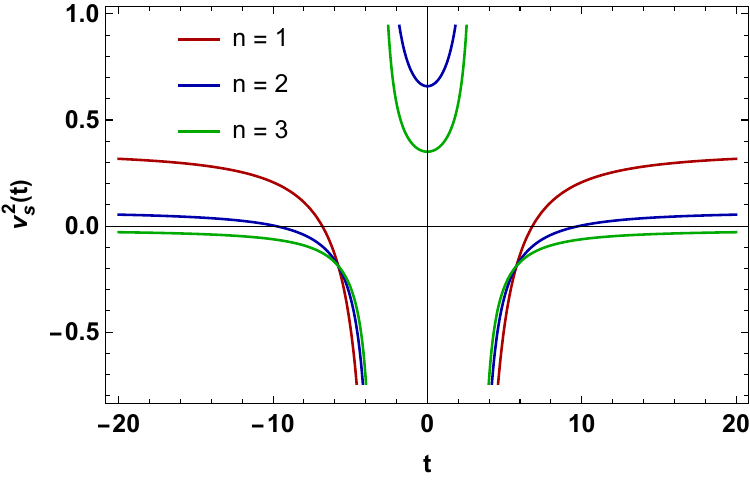}}
\caption{Temporal evolution of squared sound speed  in Weyl-type $f(Q,T)$ gravity.}
\label{F_vs}
\end{figure}

In Fig. \ref{F_vs}, the stability analysis of the Weyl-type $f(Q,T)$ gravity model is presented. The figure shows that the squared sound speed is initially positive near the bouncing point, indicating stability. However, instability emerges at a certain cosmic time, followed by a return to stability for specific values of $n$, notably $n=1$ and $n=2$. This result suggests that the proposed Weyl-type $f(Q,T)$ gravity model exhibits stability near the bouncing point. However, it is important to note that when the NEC is violated, there is a possibility of the formation of ghost fields, which can lead to dangerous instabilities at both the classical and quantum levels \cite{Tripathy}. The violation of the NEC can also raise concerns about the occurrence of superluminality, where certain physical quantities or information can travel faster than the speed of light. While efforts are made to construct models that respect the NEC and avoid such issues, it is crucial to carefully analyze and address these potential problems to ensure the validity and robustness of the proposed cosmological model.

\section{Conclusions}
\label{sec4}

In recent decades, cosmologists have grappled with the task of comprehending the origins and evolution of the Universe, especially given the limited availability of robust observational data. The presence of singularities and the challenges posed by the inflationary paradigm have spurred researchers to explore alternative avenues within the field of cosmology. One particularly intriguing approach that has garnered considerable attention is the matter bounce cosmology. In this study, we have undertaken an investigation into the matter bounce cosmology, focusing specifically on its formulation within the framework of Weyl-type $f(Q,T)$ gravity. Our study focuses on the coupling between the non-metricity $Q$ and the trace $T$ of the energy-momentum tensor, where the non-metricity is determined by the product of the metric and the Weyl vector, given by $\overline{Q}_{\mu \alpha \beta }=2w_{\mu }g_{\alpha \beta }$. This coupling is significant as it influences the covariant divergence of the metric tensor, thereby affecting the geometrical properties of the theory. Consequently, the Weyl vector and metric tensor play a crucial role in shaping the geometric features and dynamics of the model under consideration. We have chosen a straightforward functional form of $f(Q,T)=\alpha Q+\frac{\beta }{6\kappa ^2}T$, where $\alpha$ and $\beta$ are model parameters. This choice is motivated by previous research and provides a basis for our investigation \cite{Weyl1,Weyl3}. To determine the specific behavior of the bouncing cosmology, we employed the reconstruction method by parametrizing the scale factor associated with the bounce \cite{Odintsov/2016}. 

In addition, we have examined the behavior of the physical parameter, which provides insights into the matter bounce cosmology in our cosmological framework. Fig. \ref{F_a} illustrates the behavior of the scale factor of the Universe at the bouncing point $t_b=0$ for a given value of $\rho_c=0.12$. It can be observed that the scale factor reaches a non-zero minimum value of $a(t)=1$ at the bouncing point. The expansion rate of the Universe, represented by the Hubble parameter $H(t)$, proves to be a valuable tool in understanding the dynamics of bouncing cosmology. From Fig. \ref{F_H}, the Hubble parameter exhibits a contracting phase ($H(t) < 0$) before the bounce, followed by an expanding phase ($H(t) > 0$) after the bounce occurs at $t_b=0$. Also, Fig. \ref{F_q} displays the behavior of the deceleration parameter as a function of cosmic time $t$, both before and after the bouncing point. It can be observed that the deceleration parameter exhibits a negative behavior in both these time intervals. Based on the observations from Fig. \ref{F_EoS}, it can be concluded that the obtained model exhibits quintom behavior. The EoS parameter $\omega$ associated with the matter content of the Universe undergoes a significant transition from $\omega<-1$ to $\omega>-1$ in the vicinity of the bouncing point located at $t_b=0$. This transition signifies a change in the nature of the matter content, with a shift from a phase characterized by phantom-like behavior ($\omega<-1$) to a phase resembling quintessence ($\omega>-1$) \cite{Cai/2007,Cai/2009}. The occurrence of this transition in the EoS parameter near the bouncing point highlights the unique nature of the proposed model and its ability to describe the evolution of the Universe through distinct cosmological phases.

In the considered model, it is evident from the analysis that the NEC is violated in the vicinity of the bouncing point located at $t_b=0$. This violation of the NEC is significant as it satisfies the criteria for a bouncing cosmology, indicating the avoidance of singularities and the occurrence of a smooth transition between contracting and expanding phases. On the other hand, the violation of the SEC indicates the presence of exotic matter in the Universe. This violation suggests that the energy conditions associated with conventional matter are not fulfilled, and the presence of unconventional forms of matter or energy is required to drive the bouncing cosmology. Furthermore, based on the available data and analysis, the model exhibits stability near the bouncing point, indicating that it does not suffer from significant instabilities or fluctuations. The model's stability is an important aspect, ensuring its reliability and robustness in describing the dynamics of the Universe within the considered framework.

\section*{Acknowledgments}
This research is funded by the Science Committee of the Ministry of Science and Higher Education of the Republic of Kazakhstan (Grant No. AP19674478)

\textbf{Data availability} There are no new data associated with this
article.

\end{document}